\documentclass[12pt]{iopart}

\usepackage{iopams}
\usepackage{bm}
\begin{document}

\title[...Neutrino in background matter]
{Quantum treatment of neutrino in background matter}

\author{A I Studenikin}
\address{Department of Theoretical Physics,
Moscow State University, 119992 Moscow, Russia}
\ead{studenik@srd.sinp.msu.ru}

\begin{abstract}
Motivated by the need of elaboration of the quantum theory of the
spin light of neutrino in matter ($SL\nu$), we have studied in
more detail the exact solutions of the Dirac equation for
neutrinos moving in the background matter. These exact neutrino
wave functions form a basis for a rather powerful method of
investigation of different neutrino processes in matter, which is
similar to the Furry representation of quantum electrodynamics in
external fields. Within this method we also derive the
corresponding Dirac equation for an electron moving in matter and
consider the electromagnetic radiation ("spin light of electron in
matter", $SLe$) that can be emitted by the electron in this case.
\end{abstract}

\maketitle

\date{}
\maketitle

\section{Introduction}
The presence of matter has a considerable impact on neutrinos. The
resonant neutrino flavour oscillations in matter
\cite{L.Wol78,MikSmi85} has been proven to be the mechanism for
solving the solar neutrino problem. The resonant neutrino spin (or
spin-flavour) oscillations in matter \cite{Akh88,LimMar88} have
also important consequences in astrophysics and cosmology (see,
for instance, \cite{Raf_book}). In the presence of matter a
neutrino dispersion relation is modified
\cite{ManPRD88}-\cite{NiePRD89}, in particular, it has a minimum
at nonzero momentum \cite{ChaZiaPRD88,PanPLB91-PRD92WeiKiePRD97}.
As it was shown in \cite{ManPRD88}-\cite{NiePRD89}, the standard
result for the MSW effect can be derived using a modified Dirac
equation for the neutrino wave function with a matter potential
proportional to the density being added. The problem of a neutrino
mass generation in different media was also studied
\cite{OraSemSmoPLB89,HaxZhaPRD91} on the basis of modified Dirac
equations for neutrinos. On the same basis spontaneous
neutrino-pair creation in matter was also studied
\cite{LoePRL90}-\cite{KacPLB98KusPosPLB02} (neutrino-pair creation
in a perturbative way has been recently discussed in
\cite{Koe04092549}).

In \cite{BerVysBerSmiPLBGiuKimLeeLamPRD92,BerRosPLB94} another
important effect of the matter influence on neutrinos was
considered. It was shown that a majorana neutrino moving in matter
can decay into antineutrino with emission of the light scalar
majoron:
$\label{nu_nu_phi}
  \nu\rightarrow \tilde \nu + \phi.$
This process could exist due to the modification of the neutrino
spectrum induced by the presence of the background matter. In the
case of a high enough density this reaction is open even for very
light or massless neutrinos. The total decay width of the process
can be written as
 $ \Gamma = \frac {h^2}{8\pi} V, \ \ V=\sqrt 2 G_F \frac
{\rho_{eff}}  {m_n}$
\cite{BerVysBerSmiPLBGiuKimLeeLamPRD92,BerRosPLB94}, where $h$ is
the coupling constant of neutrinos to majorons, ${\rho_{eff}}$ is
the effective matter density  and $m_n$ is the neutron mass. It
should be noted that the total width does not depend on the
neutrino energy, meanwhile it is determined by the density of the
background matter. The presence of matter can also induce the
decay process of the light majoron into the couple of neutrinos
(or antineutrinos) \cite{BerRosPLB94}: $\phi\rightarrow 2\nu$ (or
$\phi\rightarrow 2\tilde \nu$).

In a series of our papers \cite{EgoLobStuPLB00-LobStuPLB04} the
quasi-classical treatment of the background matter influence on
neutrinos was developed. In particular, we proposed the
generalized Bargmann-Michel-Telegdi equation which we used for
description of the neutrino spin evolution in matter. Within this
approach we predicted \cite{LobStuPLB03LobStuPLB04} a new type of
electromagnetic radiation that can be produced by the Dirac
neutrino with nonzero magnetic moment \cite{FujShr80} in the
background matter (for a recent discussion on the neutrino
electromagnetic properties, see \cite{DvoStuPRD04JETP04}). We have
termed this radiation the "spin light of neutrino" ($SL\nu$) in
matter. The quasi-classical theory of $SL\nu$ has been developed
in \cite{LobStuPLB03LobStuPLB04,DvoGriStuIJMP05}. However, $SL\nu$
is a quantum phenomenon by its nature,  and as it has become clear
the quantum treatment of this process should be elaborated
\cite{StuTerPLB05}-\cite{GriStuTerPLB_05}.

In this paper we present in more detail the quantum treatment of neutrinos in the presence of a
dense background matter which implies the use of neutrino energy spectra and exact solutions of
modified Dirac equation for neutrino in matter. The approach developed is similar to the Furry
representation of quantum electrodynamics, widely used for description of particle interactions in
the presence of external electromagnetic fields. This approach establishes a basis for an effective
method for investigations of different phenomena which can arise  when neutrinos move in dense
astrophysical or cosmological media. We show below how the developed quantum description of the
neutrino in the presence of the background matter can be used for elaboration of the quantum theory
of the spin light of neutrino. Within this method applied to an electron, we also derive the
corresponding Dirac equation for an electron moving in matter and consider the electromagnetic
radiation (which we termed the "spin light of electron in matter", $SLe$) that can be emitted in
this case.

\section{Neutrino quantum states in matter}

 In \cite{StuTerPLB05}  (see also
\cite{StuTerQUARKS_04_0410296}-\cite{GriStuTerPLB_05}) we derived
the modified Dirac equation for the neutrino wave function exactly
accounting for the neutrino interaction with matter. Let us
consider the case of matter composed of electrons, neutrons, and
protons and also suppose that the neutrino interaction with
background particles is given by the standard model supplied with
the singlet right-handed neutrino. The corresponding addition to
the effective interaction Lagrangian is given by
\begin{equation}\label{Lag_f}
\Delta L_{eff}=-f^\mu \Big(\bar \nu \gamma_\mu {1+\gamma^5 \over
2} \nu \Big), \ \ f^\mu={\sqrt2 G_F }\sum\limits_{f=e,p,n}
j^{\mu}_{f}q^{(1)}_{f}+\lambda^{\mu}_{f}q^{(2)}_{f},
\end{equation}
where
\begin{equation}\label{q_f}
 \fl q^{(1)}_{f}=
(I_{3L}^{(f)}-2Q^{(f)}\sin^{2}\theta_{W}+\delta_{ef}), \
q^{(2)}_{f}=-(I_{3L}^{(f)}+\delta_{ef}), \  \delta_{ef}=\left\{
\begin{tabular}{l l}
1 & for {\it f=e}, \\
0 & for {\it f=n, p}. \\
\end{tabular}
\right.
\end{equation}
Here $I_{3L}^{(f)}$ and $Q^{(f)}$ are, respectively, the values of
the isospin third components and electric charges of the particles
of matter ($f=e,n,p$). The corresponding currents $j_{f}^{\mu}$
and polarization vectors $\lambda_{f}^{\mu}$ are
\begin{equation}
j_{f}^\mu=(n_f,n_f{\bf v}_f), \label{j}
\ \ \ \lambda_f^{\mu} =\Bigg(n_f ({\bm \zeta}_f {\bf v}_f ), n_f
{\bm \zeta}_f \sqrt{1-v_f^2}+ {{n_f {\bf v}_f ({\bm \zeta}_f {\bf
v}_f )} \over {1+\sqrt{1- v_f^2}}}\Bigg),
\end{equation}
$\theta _{W}$ is the Weinberg angle. In the above formulas
(\ref{j}), $n_f$, ${\bf v}_f$ and ${\bm \zeta}_f \ (0\leq |{\bm
\zeta}_f |^2 \leq 1)$ stand, respectively, for the invariant
number densities, average speeds and polarization vectors of the
matter components. A detailed discussion on the meaning of these
characteristics can be found in \cite{EgoLobStuPLB00-LobStuPLB04}.
Using the standard model Lagrangian with the extra term
(\ref{Lag_f}), we derive the modified Dirac equation for the
neutrino wave function in matter:
\begin{equation}\label{new} \Big\{
i\gamma_{\mu}\partial^{\mu}-\frac{1}{2}
\gamma_{\mu}(1+\gamma_{5})f^{\mu}-m \Big\}\Psi(x)=0.
\end{equation}
This is the most general form of the equation for the neutrino
wave function in which the effective potential
$V_{\mu}=\frac{1}{2}(1+\gamma_{5})f_{\mu}$ includes both the
neutral and charged current interactions of the neutrino with the
background particles and which could also account for effects  of
matter motion and polarization. It should be mentioned that other
modifications of the Dirac equation were previously used in
\cite{ManPRD88}-\cite{HaxZhaPRD91} for studies of the neutrino
dispersion relations, neutrino mass generation and neutrino
oscillations in the presence of matter. Note that the
corresponding quantum wave equation for a Majorana neutrino can be
obtained from (\ref{new}) via the substitution $1+\gamma _5
\rightarrow 2\gamma _5$ (see also
\cite{GriStuTerNANP_PAN06,PivStuPOS05}).

 For several important cases the equation
(\ref{new}) can be solved exactly. In particular, in the case of
the neutrino motion in matter at rest we get \cite{StuTerPLB05}
\begin{equation}\label{wave_function}
\Psi_{\varepsilon, {\bf p},s}({\bf r},t)=\frac{e^{-i(
E_{\varepsilon}t-{\bf p}{\bf r})}}{2L^{\frac{3}{2}}}
\left(%
\begin{array}{c}{\sqrt{1+ \frac{m}{ E_{\varepsilon}-\alpha m}}}
\ \sqrt{1+s\frac{p_{3}}{p}}
\\
{s \sqrt{1+ \frac{m}{ E_{\varepsilon}-\alpha m}}} \
\sqrt{1-s\frac{p_{3}}{p}}\ \ e^{i\delta}
\\
{  s\varepsilon\sqrt{1- \frac{m}{ E_{\varepsilon}-\alpha m}}} \
\sqrt{1+s\frac{p_{3}}{p}}
\\
{\varepsilon\sqrt{1- \frac{m}{ E_{\varepsilon}-\alpha m}}} \ \
\sqrt{1-s\frac{p_{3}}{p}}\ e^{i\delta}
\end{array}%
\right),
\end{equation}
where neutrino energy is
\begin{equation}\label{Energy}
  E_{\varepsilon}=\varepsilon{\sqrt{{\bf p}^{2}\Big(1-s\alpha \frac{m}{p}\Big)^{2}
  +m^2} +\alpha m}
\end{equation}
(here $p$ and $m$ are the neutrino momentum and mass, and
$\delta=\arctan{p_2/p_1}$). The matter density parameter $\alpha$
depends on the  type of neutrino and  matter composition. For
neutrinos of different flavours and matter composed of electrons,
neutrons and protons we get
\begin{equation}\label{alpha}
  \alpha_{\nu_e,\nu_\mu,\nu_\tau}=
  \frac{1}{2\sqrt{2}}\frac{G_F}{m}\Big(n_e(4\sin^2 \theta
_W+\varrho)+n_p(1-4\sin^2 \theta _W)-n_n\Big),
\end{equation}
where $\varrho=1$ for the electron neutrino $\nu_e$, and
$\varrho=-1$ in the case of the muon $\nu_\mu$ or tau $\nu_\tau$
neutrinos. The value $\varepsilon=\pm 1$ splits the solutions into
two classes that in the limit of vanishing matter density,
$\alpha\rightarrow 0$, represent the positive- and
negative-frequency solutions of the Dirac equation in vacuum.

Let us discuss in some detail the obtained neutrino energy
spectrum in matter (\ref{Energy}). The neutrino dispersion
relations in matter exhibits a very fascinating feature (see also
\cite{ChaZiaPRD88,PanPLB91-PRD92WeiKiePRD97}): the neutrino energy
may has a minimum at non-zero momentum.
It may also happen that the neutrino group and phase velocities
are  oppositely directed.  For the fixed value of the neutrino
momentum $\bf p$, there are four different energy values which are
characterized by four different combinations of the helicity
$s=\pm 1$ and energy sign $\varepsilon =\pm 1$:
\begin{equation}\label{Energy_nu}
 \fl  E^{s=+1}={\sqrt{{\bf p}^{2}\Big(1-\alpha \frac{m}{p}\Big)^{2}
  +m^2} +\alpha m}, \ \ \
 E^{s=-1}={\sqrt{{\bf p}^{2}\Big(1+\alpha \frac{m}{p}\Big)^{2}
  +m^2} +\alpha m},
\end{equation}
\begin{equation}\label{Energy_anti_nu}
 \fl {\tilde E}^{s=+1}={\sqrt{{\bf p}^{2}
  \Big(1-\alpha \frac{m}{p}\Big)^{2}
  +m^2} -\alpha m}, \ \ \
  {\tilde E}^{s=-1}={\sqrt{{\bf p}^{2}
  \Big(1+\alpha \frac{m}{p}\Big)^{2}
  +m^2} -\alpha m}.
\end{equation}
The first pair of energies is for the positive and
negative-helicity neutrino states, and the second pair is for the
corresponding antineutrino states. From the obtained energy
spectrum we conclude \cite{StuTerPLB05} that several interesting
phenomena may appear when neutrino moves in the vicinity of the
interface between media with different densities and, in
particular, between vacuum and medium with finite density. As it
follows from (\ref{Energy_nu}) and (\ref{Energy_anti_nu}), the
forbidden  energy zone for neutrino and antineutrino states in
vacuum $-m<E<m$ is shifted to $\alpha m- m< E<\alpha m + m$ in the
presence of matter. Consequently, a neutrino moving in vacuum with
the energy in the range $\alpha m- m< E<\alpha m + m$ can not
penetrate into the medium because the corresponding energy range
is forbidden in matter, and thus it will be reflected from the
interface. The analogous effect of reflection can appear at the
interface between media with different densities. Moreover, under
certain conditions it may happen that the neutrino (or
antineutrino) can not escape from a dense medium into the vacuum,
so that the neutrino trapping effect in a dense object may exist
(see also\cite{LoePRL90}). The similar analysis of the neutrino
energy spectra in matter
  shows that there may
also exist the processes of the neutrino-antineutrino pair
annihilation \cite{ChaZiaPRD88,PanPLB91-PRD92WeiKiePRD97} as well
as the spontaneous neutrino pair creation
\cite{KacPLB98KusPosPLB02} in media.

Note that the obtained energy spectrum of the flavour neutrinos
with different helicities in the presence of matter provides the
correct description  of the neutrino flavour and spin oscillations
resonance amplification in matter. Indeed, as it follows from
(\ref{Energy}) and (\ref{alpha}), the energies of the relativistic
active electron and muon neutrinos are
\begin{equation}E_{\nu_e, \nu_{\mu}}^{s=-1}\approx E_0
+ 2\alpha_{\nu_e, \nu_{\mu}} m_{\nu_e, \nu_{\mu}},
\end{equation}
where $m_{\nu_{e}}$ and $m_{\nu_{\mu}}$ are the masses of the
electron and muon neutrinos. The energy difference for the two
flavour neutrinos is
\begin{equation}\Delta E =
E_{\nu_e}^{s=-1}-E_{\nu_{\mu}}^{s=-1} = \sqrt2 G_F n_e,
\end{equation}
that provides the standard expression for the MSW effect
\cite{L.Wol78, MikSmi85}. If one considers the neutrino
spin-flavour oscillations $\nu_{e_L}\leftrightarrow \nu_{\mu_R}$
then the correspondent difference is
\begin{equation}\Delta E =
E_{\nu_e}^{s=-1}-E_{\nu_{\mu}}^{s=+1} = \sqrt2 G_F
\big(n_e-{1\over 2}n_n\big),
\end{equation}
in agreement with \cite{Akh88,LimMar88}.

\section{Quantum theory of neutrino spin light in matter}

The $SL\nu$ in matter originates from the quantum electromagnetic
transition between the two neutrino "energy levels" \ in matter.
Therefore, the most consistent theory of this phenomena can be
obtained within the quantum treatment
\cite{StuTerPLB05}-\cite{GriStuTerPLB_05} based on the use of the
modified Dirac equation (\ref{new}) exact solutions. Note that the
radiation, which is similar to $SL\nu$ in matter, can be also
emitted by a neutrino moving in the presence of a magnetic field
\cite{BorZhuTer88}.

The transition amplitude between the neutrino initial $\psi_{i}$
and final $\psi_{f}$ states can be written in the following form:
\begin{equation}\label{amplitude}
\fl \ \ \ \  S_{f i}=-\mu \sqrt{4\pi}\int d^{4} x {\bar
\psi}_{f}(x)
  ({\hat {\bf \Gamma}}{\bf e}^{*})\frac{e^{ikx}}{\sqrt{2\omega L^{3}}}
   \psi_{i}(x),\ \ \ \
   \hat {\bf \Gamma}=i\omega\big\{\big[{\bf \Sigma} \times
  {\bm \kappa}\big]+i\gamma^{5}{\bf \Sigma}\big\},
\end{equation}
where $\mu$ is the neutrino magnetic moment ,
$k^{\mu}=(\omega,{\bf k})$ and ${\bf e}^{*}$ are the momentum and
polarization of the emitted photon, and ${\bm \kappa}={\bf
k}/{\omega}$ is the unit vector in the direction of the photon
radiation. Let us consider the case of matter composed of
electrons (for simplicity we neglect possible effects of matter
motion and polarization). Then from the law of energy-momentum
conservation,
\begin{equation}\label{e_m_con}
    E=E'+\omega, \ \ \
    {\bf p}={\bf p'}+{\bf k},
    \end{equation}
it follows that the $SL\nu$ radiation arises in the transition of
the initial neutrino state with negative helicity
 $s_{i}=-1$ to the final state with positive helicity $s_{f}=+1$.
For the emitted photon energy we get
\begin{equation}\label{omega1}
\omega =\frac{2\alpha mp\left[ (E-\alpha m)-\left( p+\alpha
m\right) \cos \theta \right] }{\left( E-\alpha m-p\cos \theta
\right) ^{2}-\left( \alpha m\right) ^{2}},
\end{equation}
where the angle $\theta$ gives the direction of the radiation in
respect to the initial neutrino momentum $\bf p$. For the $SL\nu$
radiation rate and total power we get, respectively,
\begin{equation}
\label{Gamma}
 \Gamma =\mu ^2
 \int_{0}^{\pi }\frac{\omega ^{3}}{1+\tilde\beta ^{\prime
}y}S\sin \theta d\theta,   \ \ \
I=\mu ^2\int_{0}^{\pi }\frac{\omega ^{4}}{1+\tilde\beta
^{\prime}y}S\sin \theta d\theta,
\end{equation}
 where
\begin{equation}\label{S}
S=(\tilde\beta \tilde\beta ^{\prime }+1)(1-y\cos \theta
)-(\tilde\beta +\tilde\beta ^{\prime }) (\cos \theta -y),
\end{equation}
\begin{equation}\label{beta}
\tilde \beta =\frac{p+\alpha m}{E-\alpha m}, \ \ \tilde \beta
^{\prime }=\frac{p^{\prime }-\alpha m}{E^{\prime }-\alpha m}, \ \
\ E^{\prime }=E-\omega , \ \ \ p^{\prime }=K\omega -p,
\end{equation}
\begin{equation}
y=\frac{\omega -p\cos \theta }{p^{\prime }}, \ \ K=\frac{E-\alpha
m-p\cos \theta }{\alpha m}.
\end{equation}
In the relativistic neutrino momentum case, $p\gg m$, and for
different values of the matter density parameter  $\alpha$ from
(\ref{Gamma}) and we have the following limiting values
\cite{StuTerPLB05}-\cite{GriStuTerPLB_05}:
\begin{equation}\label{p_gg}
\fl \Gamma = \left\{
  \begin{tabular}{c}
  \ $\frac{64}{3} \mu ^2 \alpha ^3 p^2 m,$ \\
  \ $4 \mu ^2 \alpha ^2 m^2 p$, \\
  \ $4 \mu ^2 \alpha ^3 m^3$,
  \end{tabular}
\right. \ \ I= \left\{
  \begin{tabular}{cc}
  \ $\frac{128}{3}\mu ^{2}\alpha ^{4}p^{4},$ &
  \ \ \ for {$\alpha \ll \frac{m}{p},$ } \\
  \ $\frac{4}{3} \mu ^2 \alpha ^2 m^2 p^2$, & \ \ \ \ \ \ \ \
  { for
  $ \frac{m}{p} \ll \alpha \ll \frac{p}{m},$} \\
\ $4 \mu ^2 \alpha ^4 m^4$, & \ { for
  $ \alpha \gg \frac{p}{m}. $}
  \end{tabular}
\right.
\end{equation}
In the opposite case of "non-relativistic" neutrinos, $p\ll m$, we
get:
\begin{equation}\label{p_ll} \fl \Gamma = \left\{
  \begin{tabular}{c}
  \ $\frac{64}{3} \mu ^2 \alpha ^3 p^3,$ \\
  \ $\frac{512}{5} \mu ^2 \alpha ^6 p^3$,\\
  \ $4 \mu ^2 \alpha ^3 m^3$,
  \end{tabular}
  \ \ \
I = \left\{
  \begin{tabular}{cc}
  \ $\frac{128}{3} \mu ^2 \alpha ^4 p^4,$ &
  \ \ \ \ \ for {$\alpha \ll 1,$ } \\
  \ $\frac{1024}{3} \mu ^2 \alpha ^8 p^4$, & \ \ \ \ \ \ \ \ \ \
  {
  for
  $ 1 \ll \alpha \ll \frac{m}{p},$} \\
  \ $4 \mu ^2 \alpha ^4 m^4$, & \ \ \ \ { for
  $ \alpha \gg \frac{m}{p}. $}
  \end{tabular}
\right. \right.
\end{equation}
It can be seen that in the case of a very dense matter the values
of the rate and total power are mainly determined by the density.
Note that the obtained above results in the case of small
densities are in agreement with the studies of the neutrino spin
light performed on the basis of the quasi-classical approach
\cite{EgoLobStuPLB00-LobStuPLB04}. The $SL\nu$ characteristics in
the case of matter with "moderate" \ densities (the second lines
of (\ref{p_gg})) were also obtained in \cite{LobPLB05}.

One can estimate the average emitted photon energy $\left\langle \omega\right\rangle =
{I}/{\Gamma}$ with the use of the obtained above values of the rate and total power (\ref{p_gg})
and (\ref{p_ll}) for different matter densities. In the two case ($p\gg m$ and $p\ll m$), we get,
respectively,
\begin{equation}\label{overage_omega}
\fl \left\langle \omega\right\rangle \simeq \left\{
  \begin{tabular}{cc}
  \ $2\alpha \frac{p^{2}}{m},$ &
  for {$\alpha \ll \frac{m}{p},$ } \\
  \ $\frac{1}{3} p $, & { for
  $ \frac{m}{p} \ll \alpha \ll \frac{p}{m},$} \\
\ $\alpha m$, & \ { for
  $ \alpha \gg \frac{p}{m} $},
  \end{tabular}
\right.
\left\langle \omega\right\rangle \simeq\left\{
  \begin{tabular}{cc}
  \ $2 \alpha p ,$ &
  for {$\alpha \ll 1,$ } \\
  \ $\frac{10}{3}\alpha^2 p$, &
{ for
  $ 1 \ll \alpha \ll \frac{m}{p},$} \\
\ $\alpha  m $, &  { for
  $ \alpha \gg \frac{m}{p}. $}
  \end{tabular}
\right.
\end{equation}

To summaries the main properties of the spin light of neutrino in
matter, we should like to point out that this phenomenon arises
due to neutrino energy dependence in matter on the neutrino
helicity state. In media characterized by the positive values of
the parameter $\alpha$, the negative-helicity neutrinos (the
left-chiral relativistic neutrinos) are converted into the
positive-helicity neutrinos (the right-chiral relativistic
neutrinos) in the process under consideration. Thus, the neutrino
self-polarization effect can appear (see also
\cite{LobStuPLB03LobStuPLB04}). From the above estimations for the
emitted photon energies it follows that for the relativistic
neutrinos moving in very dense matter the $SL\nu$ can be regarded
as an effective mechanism for production of the gamma-rays. In
this concern we propose (see also \cite{GriStuTerPLB_05}) that
this mechanism can be relevant to the fireball model of GRBs and
to the description of the gamma-rays from the collapses or
coalescence processes of neutron stars, as well as for the
radiation originated during a neutron star being "eaten up" by the
black hole at the center of our Galaxy.

\section{Conclusion}

Motivated by the need of elaboration of the quantum theory of the
spin light of neutrino in matter, we have studied in detail the
exact solutions of the Dirac equation for neutrinos moving in the
background matter. In application of the obtained exact solutions
to the $SL\nu$ process we have clearly demonstrated that it is
possible to develop a rather powerful method of investigation of
different neutrino processes in matter, which is similar to the
Furry representation of quantum electrodynamics in external
electromagnetic fields. Note that in addition to the obtained
neutrino wave functions in matter it is also important to get the
explicit expression for the corresponding neutrino Green function
\cite{PivStuPOS05}.

To conclude, we should like to argue that the developed quantum
approach to description of neutrinos in the presence of background
matter can be also applied to electrons (and other particles)
moving in matter. Let us consider an electron propagating in
electrically neutral matter composed of neutrons, electrons and
protons. This situation can be realized, for instance, when
electrons move in matter of a neutron star. We suppose that there
is a macroscopic amount of the background particles in the scale
of an electron de Broglie wave length. Then the addition to the
electron effective interaction Lagrangian is
\begin{equation}\label{Lag_f_e}
\Delta L^{(e)}_{eff}=-f^\mu \Big(\bar e \gamma_\mu
{1-4\sin^{2}\theta_{W}+\gamma^5 \over 2} e \Big),
\end{equation}
where $f^{\mu}$ is determined by (\ref{Lag_f}) and (\ref{q_f})
with $\delta_{ef}=0$. The modified Dirac equation for the electron
wave function in matter is
\begin{equation}\label{new_e} \Big\{
i\gamma_{\mu}\partial^{\mu}-\frac{1}{2}
\gamma_{\mu}(1-4\sin^{2}\theta_{W}+\gamma_{5}){\tilde f}^{\mu}-m_e
\Big\}\Psi_{e}(x)=0,
\end{equation}
where
\begin{equation}
{\tilde f}^{\mu}=-f^\mu=\frac{G_F}{\sqrt
2}(j^{\mu}_n-\lambda^{\mu}_n).
\end{equation}
The corresponding electron energy spectrum in the case of
unpolarized matter at rest is given by
\begin{equation}\label{Energy_e}
  E_{\varepsilon}^{(e)}=
  \varepsilon \sqrt{{{\bf p}_e}^{2}\Big(1-s_e\alpha_n
  \frac{m_e}{p_e}\Big)^{2}
  +{m_e}^2} +c{\alpha}_n m_e, \ \ \alpha_n=\frac{1}{2\sqrt{2}}
  {G_F}\frac{n_n}{m_e},
\end{equation}
where $c=1-4\sin^2 \theta_W$ and  the notations for the electron
mass, momentum, helicity and sign of energy are similar to those
used in Section 2 for the case of neutrino. The exact solutions of
this equation open a new method for investigation of different
quantum processes which can appear when electrons propagate in
matter. On this basis, we predict a mechanism of the
electromagnetic radiation by an electron moving in matter which we
term the "spin light of electron in matter" ($SL\e$). Note that
the similar term ("spin light of electron in magnetic field") was
used previously for the magnetic moment dependent contribution to
the radiation of an electron in a magnetic field
\cite{BorTerBag95}. The $SL\e$ photon energy, obtained from the
energy conservation law, is given by
\begin{equation}\label{omega1_e}
\omega_{SLe} =\frac{2\alpha_n m_ep_e\left[ (E^{(e)}-c\alpha_n
m_e)-\left( p_e+\alpha_n m_e\right) \cos \theta_{SLe} \right]
}{\left( E^{(e)}-c\alpha_n m_e-p_e\cos \theta_{SLe} \right)
^{2}-\left( \alpha_n m_e\right) ^{2}},
\end{equation}
where the angle $\theta_{SLe}$ gives the direction of the
radiation in respect to the initial electron momentum ${\bf p}_e$.
In the case of relativistic electrons and small values of the
matter density parameter $\alpha_n$ the photon energy is
\begin{equation}\label{omega_2}
    \omega_{SLe}=
    \frac {\beta_e}{1-\beta_e \cos
    \theta_{SLe}}\omega_0,\ \ \
\omega_0= \frac {G_{F}} {\sqrt{2}}n_n\beta_e,
\end{equation}
here $\beta_e$ is the electron speed in vacuum. From this
expressions we conclude that for the relativistic electrons the
energy range of the $SLe$ may even extend up to energies peculiar
to the spectrum of gamma-rays. We also predict the existence of
the electron-spin polarization effect in this process. Finally,
from the order-of-magnitude estimation, we expect that the ratio
of rates of the $SL\e$ and the $SL\nu$ in matter is
\begin{equation}
R=\frac {\Gamma_{SLe}}{\Gamma_{SL\nu}}\sim \frac {e^2}{\omega^2
\mu^2},
\end{equation}
that gives $R \sim 10^{18}$ for the radiation in the range of
gamma-rays, $\omega \sim 5 \ MeV$, and for the neutrino magnetic
moment $\mu \sim 10^{-10}\mu_0$. Thus, we expect that in certain
cases the $SLe$ in matter would be more effective than the
$SL\nu$.  For the detailed study of the $SLe$ in matter see
\cite{ShiStuTerTro96}.

 I would like to thank Emilio Elizalde and Sergei Odintsov for
the invitation to participate to the Seventh Workshop on Quantum
Field Theory under the Influence of External Conditions. I am also
thankful to all of the organizers of this workshop for their
hospitality.

\end{document}